\documentclass[apj,numberedappendix]{emulateapj}

\shorttitle{COMPOSITION AND NOVA IGNITIONS}
\shortauthors{SHEN \& BILDSTEN}

\newcommand{\be}{\begin{eqnarray}}
\newcommand{\ee}{\end{eqnarray}}
\newcommand{\lp}{\left(}
\newcommand{\rp}{\right)}

\newcommand{\E}[1]{\times10^{#1}}
\newcommand{\smpy}{M_\odot \ {\rm yr}^{-1}}

\begin{document}


\title{The Effect of Composition on Nova Ignitions}

\author{Ken J. Shen\altaffilmark{1,2} \& Lars Bildsten\altaffilmark{1,3}}
\altaffiltext{1}{Department of Physics, Broida Hall, University of California, Santa Barbara, CA 93106; kenshen@physics.ucsb.edu.}
\altaffiltext{2}{Max-Planck Institut f\"{u}r Astrophysik, Karl-Schwarzschild-Str. 1, 85748 Garching, Germany.}
\altaffiltext{3}{Kavli Institute for Theoretical Physics, Kohn Hall, University of California, Santa Barbara, CA 93106; bildsten@kitp.ucsb.edu.}


\begin{abstract}

The accretion of hydrogen-rich matter onto C/O and O/Ne white dwarfs in binary systems leads to unstable thermonuclear ignition of the accreted envelope, triggering a convective thermonuclear runaway and a subsequent classical, recurrent, or symbiotic nova.  Prompted by uncertainties in the composition at the base of the accreted envelope at the onset of convection, as well as the range of abundances detected in nova ejecta, we examine the effects of varying the composition of the accreted material.  For high accretion rates and carbon mass fractions $< 2\E{-3}$, we find that carbon, which is usually assumed to trigger the runaway via proton captures, is instead depleted and converted to $^{14}$N.  Additionally, we quantify the importance of $^3$He, finding that convection is triggered by $^3$He$+^3$He reactions for $^3$He mass fractions $> 2\E{-3}$.  These different triggering mechanisms, which occur for critical abundances relevant to many nova systems, alter the amount of mass that is accreted prior to a nova, causing the nova rate to depend on accreted composition.  Upcoming deep optical surveys such as Pan-STARRS-1, Pan-STARRS-4, and the Large Synoptic Survey Telescope may allow us to detect the dependence of nova rates on accreted composition.  Furthermore, the burning and depletion of $^3$He with a mass fraction of $10^{-3}$, which is lower than necessary for triggering convection, still has an observable effect, resulting in a pre-outburst brightening in disk quiescence to $> L_\odot$ and an increase in effective temperature to $6.5\E{4}$ K for a $1.0 \ M_\odot$ white dwarf accreting at $10^{-8} \ \smpy$.

\end{abstract}



\section{Introduction}

White dwarfs (WDs) in cataclysmic variable (CV) and symbiotic binary systems accrete H-rich material from main sequence and red giant donors, respectively, with accretion rates of $ \dot{M} = 10^{-11} - 10^{-7} \ \smpy $ \citep{war95}.  As the accreted envelope gains mass, compression of the material at the base of the layer leads to a temperature increase, eventually triggering H-burning.  Depending on $\dot{M}$, the WD mass, and the composition of the accreted material, the eventual outcome will either be steady thermally stable H-burning \citep{sien75,sien80,pac83,fuji82b,sb07,nom07}, or a convective thermonuclear runaway \citep{starr72,fuji82a}, observable as a classical, recurrent, or symbiotic nova when the convective zone nears the WD photosphere.  (Hereafter, we will refer to the onset of convection and the ignition of the nova interchangeably.)

It was first recognized from numerical simulations \citep[e.g.,][]{starr72} that the mass fraction of CNO isotopes, $Z_{\rm CNO}$, in the burning layers during a nova outburst must be $10-50$ times higher than the solar value in order to eject sufficient mass and produce a light-curve similar to those observed for fast novae.  Abundance measurements confirm this metal enhancement in the ejecta of fast novae (for observational summaries, see \citealt{gehrz98} and \citealt{hk06}).  This degree of enrichment cannot be explained by accretion from an evolved donor or nucleosynthesis during the nova outburst \citep[e.g.,][]{tl86}, thus some mechanism of core dredge-up and mixing with the envelope must be invoked.

There is on-going debate over the effectiveness of proposed mixing mechanisms, which we detail in \S\ref{sec:motiv}.  In particular, it is unclear whether convection is initiated above or below the core-envelope interface in C-poor or C-rich material.  With this in mind, our work examines the effect of the accreted composition on the pre-ignition characteristics of nova systems, assuming no CNO enrichment prior to convection.  We calculate ignition masses and pre-ignition luminosities for models with a large range of metallicities centered around solar ($ Z = 0.1-5.0 \ Z_\odot $), non-zero $^3$He mass fractions ($ X_3 = 10^{-4} - 0.005 $), high accretion rates ($ \dot{M} = 10^{-9} - 3\E{-7} \ \smpy $), and a mass range of $M=0.6-1.35 \ M_\odot$.  Note that the eventual outcome for the models with a combination of high accretion rate and low WD mass is steady and stable H-burning \citep{sien75,sien80,pac83,fuji82b,sb07,nom07}, so we omit these models from our results.

In \S\ref{sec:analytics} and \S\ref{sec:3heeffect}, we use analytic approximations to motivate the more exact numerical study described in \S\ref{sec:numerics}.  Below a critical accreted carbon mass fraction $X_{12} \simeq 2\E{-3}$, which is coincidentally near the solar value, we find that $^{12}$C is depleted and converted to $^{14}$N prior to unstable ignition, so that convection is triggered by proton captures onto $^{14}$N.  We also find that convection is triggered by $^3$He$+^3$He reactions for $^3$He mass fractions $X_3 \gtrsim 2\E{-3}$.  These different nova triggers change the pre-outburst luminosity of the nova system, which we detail in \S\ref{sec:obs}.  The ignition mass is also affected by the triggering mechanism, which results in a previously unconsidered dependence of galactic nova rates on composition.  This effect is especially relevant given the upcoming flood of data from optical transient surveys such as Pan-STARRS-1, Pan-STARRS-4, and the Large Synoptic Survey Telescope, which will measure nova rates in external galaxies with greater accuracy than available in current data.  We speculate on the observational consequences and summarize our work in \S\ref{sec:conc}.


\section{Motivation for our study and Justification of Assumptions}
\label{sec:motiv}

Core-envelope mixing models differ on the mechanism by which core material is brought into the accreted layer.  Chemical diffusion \citep{pk84,kp85,pk95,yaron05} and shear mixing caused by differential rotation of the accreted material \citep{kt78,lt87,ks89,alex04} result in pre-convective penetration of a small amount of hydrogen into the underlying material and vice versa.  If the material below the layer is C-rich, $p+^{12}$C reactions trigger convection, which homogenizes the envelope and the entrained core material.  These pre-convective enrichment models differ from mechanisms in which convection is triggered above the core in accreted material whose composition is determined by that of the donor star and is thus relatively C-poor.  The introduction of core material into the envelope for these convective enrichment models is caused by the convective motion itself, either via convective overshoot into the core \citep{woos86} or shear mixing induced by the convective eddies \citep{glt97,ros01}.

As of yet, no mixing mechanism has definitively proven itself successful in explaining the enrichments of all novae.  For example, the multi-cycle diffusion studies of \cite{pk95} and \cite{yaron05} are initiated with matter accreted directly onto naked C/O cores, yet the accreting WD may be O/Ne in as many as 1/3 of all observed novae \citep{tl86,rit91,lt94,gil03}.  For these systems, diffusion would not lead to the initiation of convection below the accreted layer and subsequent core dredge-up because the underlying material is not C-rich.  The studies of \cite{lt87}, \cite{fuji88,fuji93}, and \cite{pb04} rule out any significant differential rotation between the accreted layer and the core, which casts doubt on accretion-induced shear mixing mechanisms.  \cite{kht98,kht99} find that convective overshoot and shearing do not sufficiently enrich the envelope to produce a fast nova, although possible problems with their boundary conditions are pointed out by \cite{glt05}.  Moreover, some recurrent nova, which are novae with recurrence times $\sim 30 $ yr, do not show over-abundances of metals in their ejecta \citep{wil81,wil82,war95,hk01}, possibly due to a large helium buffer above the core.

Thus, it is unclear how much the envelope will be enriched in metals prior to nova ignition for $\dot{M}>10^{-9} \ \smpy$.  Some nova studies that include the accretion phase assume that the accreting envelope is pre-enriched by the core and consists of up to 50\% core material by mass \citep[e.g.,][]{jh98,starr98}.  However, the uncertainties involved in the mixing mechanisms coupled with the lack of observed metal enrichment in recurrent novae lead us to examine the consequences of assuming no C-enrichment in the accreted envelope prior to the onset of convection.  There are several previous studies that also follow this treatment \citep{starr85,starr88,starr00,tru88,pier00,jose07}, but none sufficiently samples the full parameter space in which we are interested: \cite{starr85,starr88} and \cite{tru88} consider $Z=0.02$ accretion onto $ M \geq1.35 \ M_\odot$ WDs; the models of \cite{starr00} have metallicity $Z=10^{-3}$ or $0.02$, as motivated by novae in the Large Magellanic Cloud, with lower accretion rates $\lesssim 10^{-9} \ \smpy$ than what we study; \cite{pier00} examine accretion with 3 metallicities ($Z=0.02$, $10^{-3}$, and $10^{-4}$) onto WDs with masses $<0.68 \ M_\odot$, lower than our parameter range; and \cite{jose07}, in their study of novae in primordial binaries, have only solar and very sub-solar metallicity $1.35 \ M_\odot$ models with $Z=0.02$, $2\E{-6}$, or $10^{-7}$, with $\dot{M}=2\E{-10} \ \smpy$, which is lower than our range.  Moreover, none of these studies consider the effect of $^3$He, which can play a dominant role in triggering the nova \citep{shara80,tb04}.

In our study, we assume that the effect of chemical diffusion is negligible.  This assumption, and thus our results, are invalid if the material directly below the accreted layer is C-rich.  However, as we have described above, many nova systems exist in which the underlying material is C-poor and diffusion is indeed negligible.  Our results only apply to these systems.


\section{Carbon Depletion Prior to Unstable Ignition}
\label{sec:analytics}

If the accretion rate in a CV is lower than the minimum rate for stability, the result will be a hydrogen shell flash.  In this section, we calculate the ignition conditions for these thermonuclear novae.  Throughout this study, we make the assumption that the accreted layer is thin, with a pressure scale height at the envelope base, $h=P_b/\rho_b g$, much less than the WD core radius, $R$.  The subscript $b$ refers to the base of the accreted envelope, $g=GM/R^2$ is the gravitational acceleration, assumed constant because $h \ll R$, and $M$ is the WD core mass.  The ratio of the scale height to the WD radius for an ideal gas equation of state is the ratio of the thermal energy to gravitational energy,
\be
	\frac{h}{R} = \frac{k_B T_b / \mu m_p}{G M / R } = 10^{-2} \frac{T_7 R_9}{ M_1} ,
\ee
where $T_7$ is the base temperature in units of $10^7$ K, $ M_1 $ is the WD core mass in units of $ M_\odot $, $R_9=R/10^9$ cm, the proton mass is $m_p$, and the atomic mass per particle is $\mu=0.6$ for solar composition.  For typical ignition conditions, $T_b \simeq 2\times 10^7$ K, so the shell is very thin for the entire accretion phase.  The base pressure is independent of the temperature in the thin-shell limit and is
\be
	P_b = \frac{GM M_{\rm env}}{4\pi R^4} ,
	\label{eq:thinshell}
\ee
where $M_{\rm env}$ is the envelope mass.  This assumption will not be valid once the temperature rises during the thermonuclear runaway.  For a $0.6 \ M_\odot$ WD, $h \simeq R$ when $T_b \simeq 7\E{8} \ {\rm K}$.

We first estimate the luminosity in the accreting layer following \cite{nom82} and \cite{tb04}.  When material accretes onto the WD surface, the gravitational energy, $GM/R$, is released and radiated by the spreading boundary layer \citep{pb04} and is not carried into the star, because the thermal timescale at the photosphere for luminosities of order the accretion luminosity is far shorter than the accretion timescale.  Instead, prior to the onset of nuclear burning, the pre-ignition luminosity exiting the deep accreting layer is produced by entropy released as the material accumulates.  The entropy equation yields the compressional luminosity at the surface,
\be
	L_{\rm comp} = \dot{M} \int^0_{P'} T \frac{ ds }{ dP } \, dP ,
	\label{eq:entrpyintgrl}
\ee
where $s$ is the specific entropy, and we have neglected the term $ \left. \partial s / \partial t \right|_P$.  The lower bound, $P'$, is the depth at which the thermal time is equal to the time for which accretion has been on-going, so that the luminosity produced there has had time to make its way through the envelope.  For illustration, we assume Kramers' opacity ($\kappa \propto \rho T^{-7/2}$), ideal gas ($P \propto \rho T$), and a constant luminosity above $P'$, so that $ P(r)^2 \propto T(r)^{17/2} $.  An ideal gas has $ s = k_B \ln \lp T^{3/2} / \rho \rp / \mu m_p $, which yields $ds/dP = -7k_B/17 \mu m_p P$.  This gives
\be
	L_{\rm comp} = \frac{7}{4} \dot{M} \frac{ k_B T'}{\mu m_p} = 0.4 \ L_\odot \ \dot{M}_{-8} \left( \frac{T'}{10^7 \ {\rm K}} \right) ,
	\label{eq:complum}
\ee
where $ T' $ is the temperature at $P'$, $ \dot{M}_{-8} $ is the mass accretion rate in units of $ 10^{-8} \ \smpy$, and we have set $\mu=0.6$.  If the opacity is due to electron scattering, the pre-factor becomes $ 3/2 $ instead of $7/4$, so the exact relation is only weakly dependent on the form of radiative opacity.

The thermal time at $P'$ is $t_{\rm therm}' \equiv c_P T' M'_{\rm env} / L_{\rm comp}$, where $c_P = 5 k_B / 2 \mu m_p$ is the specific heat at constant pressure for an ideal gas, $M'_{\rm env}$ is the mass in the layer above $P'$, and we use a one-zone approximation, $ \partial L / \partial M \sim L_{\rm comp}/M' $ \citep[e.g.,][]{pac83}.  The time to accrete an envelope mass $M_{\rm env}$ is $t_{\rm acc} \equiv M_{\rm env} / \dot{M}$.  To find the depth from which luminosity is able to escape during accretion, we set the two timescales equal and use equation (\ref{eq:thinshell}), yielding $P' \simeq P_b$; i.e., most of the luminosity in the envelope comes from only the envelope itself, and so we neglect the compressional luminosity from the core (see the appendix of \citealt{tb04} for further discussion of the core's role).  Thus, the compressional luminosity is given by equation (\ref{eq:complum}), with $T'=T_b$.

Using the radiative diffusion equation with Kramers' opacity,
	\be
	\kappa = \kappa_0 \frac{\rho}{{\rm g \ cm}^{-3}} \left( \frac{T}{\rm K} \right)^{-7/2},
	\ee
where $\kappa_0 \simeq 10^{22}$ cm$^2$ g$^{-1}$ from fitting to OPAL opacities \citep{ir93,ir96} for solar composition around $T=10^7$ K and $\rho=10^3$ g cm$^{-3}$, we derive the temperature at the base of the accreting layer as a function of $\rho_b$,
\be
	T_b	= 1.37 \E{7} \ {\rm K} \lp \frac{ \dot{M}_{-8} \rho^2_3 }{ M_1 } \rp^{2/11},
	\label{eq:traj}
\ee
where $ \rho_3 = \rho_b /  10^3 $ g cm$^{-3}$.  We have assumed solar metallicity, but this result is nearly independent of composition.  The bottom of the layer follows the trajectory given by equation (\ref{eq:traj}) until nuclear burning becomes comparable to compressional heating, i.e., when $L_{\rm nuc} \sim L_{\rm comp}$, where $L_{\rm nuc}$ is the luminosity produced by nuclear burning.  For high accretion rates $\geq 10^{-9} \ \smpy$, H-burning occurs via CNO reactions when base conditions reach $T_b \simeq 2\E{7}$ K and $\rho_b \simeq 10^3$ g cm$^{-3}$ (ignoring $^3$He-burning).  If the accreting material has near-solar isotopic ratios, the most relevant isotope is $^{12}$C, since proton captures onto $^{14}$N are slower than onto $^{12}$C, and $^{16}$O does not participate in the CNO cycle at these temperatures.  Moreover, $p+p$ reactions are unimportant at $T_b \simeq 2\E{7}$ K because the lifetime of a proton with respect to self-burning is $\simeq 10$ times longer than with respect to consumption by $^{12}$C nuclei.  Thus, proton captures onto $^{12}$C will be the first non-negligible reaction.  These are quickly followed by the $\beta$-decay of $^{13}$N (with a half-life of $\tau_{1/2}=603$ s) and proton captures onto $^{13}$C ($\simeq 4$ times more rapid than onto $^{12}$C), so that we approximate the first nuclear reactions of interest as the conversion of $^{12}$C to $^{14}$N at the $^{12}$C proton capture rate.  This reaction chain releases a specific energy $ X_{12} E_{12} = 8.8\E{14} (X_{12}/10^{-3})$ erg g$^{-1}$.

Linear stability analysis \citep{fhm81} shows that nuclear burning is unstable in a constant-pressure thin shell if
\be
	\left. \frac{ \partial \epsilon_{\rm nuc} }{ \partial T } \right|_P > \left. \frac{ \partial \epsilon_{\rm cool} }{ \partial T } \right|_P ,
	\label{eq:unstab}
\ee
or $ \epsilon_{\rm nuc} \chi_{\rm nuc} > \epsilon_{\rm cool} \chi_{\rm cool} $, where $ \epsilon_{\rm nuc}$ is the nuclear energy generation rate, the one-zone approximation to the cooling rate is $ \epsilon_{\rm cool} \sim L/M_{\rm env} $, and $ \chi \equiv \left. \partial \ln \epsilon / \partial \ln T \right|_P $.  For Kramers' opacity and ideal gas, $ \chi_{\rm cool} = 17/2 $.  The cooling rate is rewritten as
\be
	\epsilon_{\rm cool} = \frac{7}{4} \frac{k_B T_b}{\mu m_p} \frac{1}{t_{\rm acc}} 
\ee
after substituting the expression for compressional luminosity (eq. [\ref{eq:complum}]).

It is often stated that the energy released by the conversion of $^{12}$C to $^{14}$N triggers the thermonuclear runaway for high $\dot{M}$ novae (e.g., \citealt{fuji82b,tb04,jose05,jh07}).  However, we show here that under some conditions this reaction will not release enough heat to trigger unstable ignition conditions before the $^{12}$C is depleted.  In this case, all available $^{12}$C converts to $^{14}$N, which will ignite later at a larger pressure and temperature.  Carbon depletion occurs when the accretion and burning timescales are comparable, with the C-burning timescale given by $ t_{12} = X_{12} E_{12} / \epsilon_{12} $, where $\epsilon_{12}$ is the rate of energy generation from conversion of $^{12}$C to $^{14}$N.  The condition for stable $^{12}$C depletion is then
\be
	X_{12} < 3\E{-3} \frac{T_b}{2\E{7} \ {\rm K}} \frac{\chi_{\rm cool}}{8.5} \frac{15}{\chi_{12}} .
	\label{eq:12Cdep}
\ee
Thus, for mass fractions below a critical value, coincidentally near the solar mass fraction of $2.2\E{-3}$, carbon will deplete before triggering a nova.  This value is certainly relevant for novae in systems with low-metallicity donors.  Furthermore, low $^{12}$C/$^{14}$N ratios can occur when mass transfer has revealed a CNO-processed core \citep[][and references therein]{schenk02}.  In these cases, the carbon mass fraction of the accreted material will be well below the solar value because proton captures onto $^{14}$N are the slow step of the CNO cycle, and thus the donor's CNO nuclei are mostly in the form of $^{14}$N, resulting in accreted carbon mass fractions $\sim 10^{-4}$ or lower \citep{schenk02}.  For high $\dot{M}$ CV systems, evolved donors such as these are likely common.  A population synthesis calculation by \cite{pod03} finds that the majority of CVs with orbital periods $P_{\rm orb}>5$ h have an evolved secondary.  Observationally, a study of UV line flux ratios of CVs both above and below the period gap \citep{gan03} concludes that as much as $10-15 \%$ of their sample might have evolved donors with anomalously low $^{12}$C/$^{14}$N abundance ratios due to CNO processing.

\begin{figure}
	\epsscale{1.0}
	\plotone{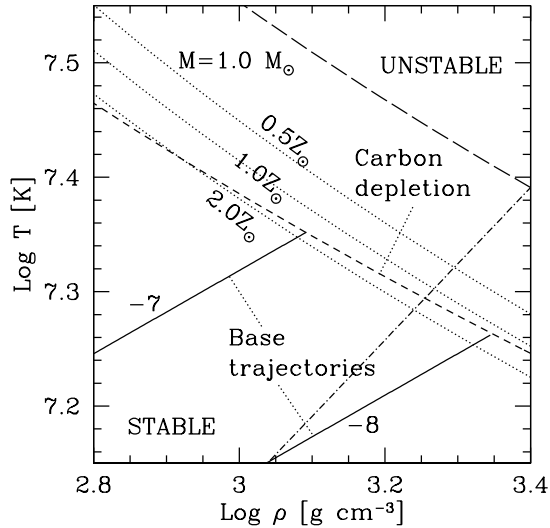}
	\epsscale{1.0}
	\caption{Analytic $^{12}$C ignition and depletion curves for a $1.0 \ M_\odot$ WD.  Dotted lines demarcate the boundary between stable $p+^{12}$C burning (\emph{lower-left region}) and unstable ignition (\emph{upper-right region}) for the given accreted metallicities.  The short-dashed line shows where $t_{\rm acc}=t_{12}$, which is independent of $\dot{M}$ and $Z$; above this line, $^{12}$C has been depleted.  The solid lines follow the trajectories at the base of envelopes accreting at $\dot{M} = 10^{-7}$ and $10^{-8} \ \smpy $; as more material accumulates, the base becomes hotter and denser, and the lines move to the upper-right, where carbon either ignites unstably or depletes.  If carbon is depleted, the base continues to heat and compress until $^{14}$N begins to burn and the full CNO is ignited (\emph{long-dashed line}).  To the left of the dashed-dotted line, the base is mildly degenerate or non-degenerate.}
	\label{fig:stabdep}
\end{figure}

Figure \ref{fig:stabdep} shows conditions for carbon depletion (\emph{short-dashed line}) and stable vs. unstable C-burning with varying accreted metallicities (\emph{dotted lines}) for a $1.0 \ M_\odot$ WD.  The region to the left of the dashed-dotted line has Fermi energy $E_F<3k_B T$, so that it is mildly degenerate or non-degenerate.  Free-free opacity dominates in the region of interest.  Also shown are the rising temperature and density at the base of accreting layers with $\dot{M} = 10^{-7}$ and $10^{-8} \ \smpy $ (\emph{solid lines}).  For mass fractions lower than near-solar, carbon depletes before it ignites unstably and is instead converted to $^{14}$N.  In these cases, the base of the layer continues to become hotter and denser until it reaches burning conditions for proton captures onto $^{14}$N (\emph{long-dashed line}).  At the temperatures and densities corresponding to these $\dot{M}$'s, the CNO cycle as a whole is thermally unstable, and ignition is inevitable.  However, because the layer must be hotter and denser to burn the $^{14}$N, more material must be accreted before ignition than in the case of carbon ignition.


\section{The Significance of $^3$He}
\label{sec:3heeffect}

The possible importance of $^3$He for novae was first studied by \cite{schatz51} in the context of a theory of novae powered by thermonuclear detonations.  \cite{mac83} and \cite{it84} noted that the presence of $^3$He can decrease the envelope mass needed for nova ignition, and \cite{shara80} and \cite{tb04} looked more closely at its role in triggering the convective shell burning phase of a nova.  Although these studies found that $^3$He can play a large role in the onset of a nova, its effects have not been quantified in detail.

For typical nova conditions, $^3$He-burning takes place via $^3$He($^3$He, 2$p$)$^4$He.  The next-fastest reaction that consumes $^3$He is $^3$He$+^4$He, which has an unscreened reaction rate that is a factor of $\sim 10^3$ slower for $X_3 = 10^{-3}$, and thus $^3$He$+^3$He is the only reaction to consider.  To gauge the importance of $^3$He-burning for nova ignitions, we compare its relevant characteristics to proton captures onto $^{12}$C, the assumed trigger for nova systems with $\dot{M} \geq 10^{-9} \ \smpy$.  The energy per $^3$He self-reaction is $12.9$ MeV, slightly larger than the $11.0$ MeV released by the reaction chain that converts $^{12}$C to $^{14}$N.  Moreover, the $^3$He unscreened reaction timescale, i.e., the e-folding lifetime for $^3$He nuclei, defined for a generic nucleus $i$ as $n_i/t_i=dn_i/dt_i$, is much shorter than that of $^{12}$C.  The ratio of timescales is
\be
	\frac{ t_{12} }{ t_3 } \sim 10^3 \ \frac{X_3}{X_{\rm H}} \ \exp \lp \frac{6.56}{T_7^{1/3}} \rp,
\ee
which is roughly a factor of 200 for $2\E{7}$ K, $X_3 =10^{-3}$, and hydrogen mass fraction $X_{\rm H}=0.75$.  Thus, for $X_3> 10^{-5}$, $^3$He nuclei will begin burning via self-reactions before $^{12}$C nuclei have a chance to capture protons.

Since the $^3$He reaction releases a similar amount of energy and has a similar temperature dependence to that of the $^{12}$C proton capture, we must also consider the possibility of $^3$He depletion prior to nova ignition.  An analysis like that of the previous section yields the same critical mass fraction $\simeq 3\E{-3}$ as in equation (\ref{eq:12Cdep}).  Again coincidentally, this critical mass fraction for $^3$He depletion is in the neighborhood of the expected value for mass-transferring binary systems.  As mass loss uncovers deeper parts of the donor star, material that has been processed by H-burning can make its way to the surface.  \cite{dm82} and \cite{it84} find that the mass fraction of accreted $^3$He can reach values as high as $4\E{-3}$ during the evolution of systems with low-mass donors, and thus the possibility exists for $^3$He to trigger a nova prior to depleting.  For this reason, \cite{tb04} included calculations with $X_3 = 0.001$ and 0.005.  For $\dot{M}>3\E{-10} \ \smpy$, they found that ignitions were triggered by $^3$He, with the ignition mass $M_{\rm ign}$ roughly decreasing by a factor of 2 when increasing $X_3$ from $0.001$ to $0.005$.


\section{Numerical Simulation}
\label{sec:numerics}

The approximations that we have made in the analytic work of \S\ref{sec:analytics} and \S\ref{sec:3heeffect} and the proximity of the critical $^{12}$C and $^3$He abundances to relevant solar and CV abundances motivate a more exact analysis.  In particular, electron degeneracy pressure, other opacities, the exchange of heat between the envelope and core, and the term $\left. \partial s / \partial t \right|_P$ that is neglected in equation (\ref{eq:entrpyintgrl}) must be included.  Moreover, the one-zone approximation of setting $ \partial L / \partial M \sim L/M_{\rm env} $ is problematic, because burning occurs in a narrow layer and is certainly not a linear function of the whole atmosphere.


\subsection{Model}

We developed a time-dependent explicit Runge-Kutta code for a one-dimensional grid of 100 zones covering a pressure range of $3\E{17} - 3\E{20}$ dyne cm$^{-2}$.  Since the base of the envelope at ignition is typically between $ 3\E{18}$ and $3\E{19}$ dyne cm$^{-2}$, this choice of zoning gives greater than a factor of 10 range in pressure above and below the region of interest.  For the $1.35 \ M_\odot$ model with $\dot{M}=10^{-9} \ \smpy$, the pressure boundaries were changed to $10^{18}-10^{21}$ dyne cm$^{-2}$ because the accreted layer reached a depth that was too close to $3\E{20}$ dyne cm$^{-2}$.  The zones are spaced logarithmically in pressure (thus there are roughly 33 zones per decade of pressure) to better resolve the accreted layer and to avoid over-resolution of the bottom zones.  The layer is spherically symmetric, appropriate for the depths of interest if there is negligible differential rotation \citep{fuji88,fuji93,pb04}, and plane-parallel, a good approximation as discussed in \S\ref{sec:analytics}.

\begin{table}
	\begin{center}
	\caption{Model Core Temperatures}
	\label{tab:coretemp}
	\begin{tabular}{|c|c|c|}
	\hline
	Mass [$M_\odot$] & $\dot{M}$ [$M_\odot \ {\rm yr}^{-1}$] & $T_c$ [K]\\
	\hline
	\hline
	0.6 & $10^{-9}$ & $8.00\E{6}$\\
	\hline
	0.6 & $10^{-8}$ & $1.04\E{7}$\\
	\hline
	0.6 & $10^{-7}$ & $1.34\E{7}$\\
	\hline
	0.6 & $3\E{-7}$ & $1.51\E{7}$\\
	\hline
	1.0 & $10^{-9}$ & $8.99\E{6}$\\
	\hline
	1.0 & $10^{-8}$ & $1.10\E{7}$\\
	\hline
	1.0 & $10^{-7}$ & $1.45\E{7}$\\
	\hline
	1.0 & $3\E{-7}$ & $1.70\E{7}$\\
	\hline
	1.2 & $10^{-9}$ & $9.18\E{6}$\\
	\hline
	1.2 & $10^{-8}$ & $1.20\E{7}$\\
	\hline
	1.2 & $10^{-7}$ & $1.80\E{7}$\\
	\hline
	1.2 & $3\E{-7}$ & $2.29\E{7}$\\
	\hline
	1.35 & $10^{-9}$ & $9.25\E{6}$\\
	\hline
	1.35 & $10^{-8}$ & $1.29\E{7}$\\
	\hline
	1.35 & $10^{-7}$ & $2.18\E{7}$\\
	\hline
	1.35 & $3\E{-7}$ & $3.00\E{7}$\\
	\hline
	\end{tabular}
	\end{center}
\end{table}

The thermal evolution of the core during CN cycles was considered in detail by \cite{tb04}, who found that heating and cooling during the nova cycle results in an equilibrium core temperature, $T_{c,\rm eq}$.\footnote{CVs spend only a short amount of time at $\dot{M} > 10^{-8} \ \smpy$ \citep{how01}, and so it is unclear if the core temperature will have time to reach equilibrium before the system has evolved appreciably \citep{ep07}.  However, as mentioned later, the results of our study are largely unaffected by the core temperature.}  Thus, for our models, the initial thermal profile prior to accretion is assumed to be a radiative-zero solution that gives a core temperature $T_c=T_{c,\rm eq}$.  The equilibrium $T_c$ depends on the accreted composition, but, in order to limit the parameters of this study, a single representative $T_c$ is used for each $M$ and $\dot{M}$ model; these are extrapolated from \cite{tb04} and Townsley (2007, priv. comm.) and are shown in Table \ref{tab:coretemp}.  Fortunately, the properties of the accreting layer are relatively insensitive to $T_c$ for these high $\dot{M}$'s.  For example, increasing $T_c$ by a factor of 2 decreases $M_{\rm ign}$ by only $< 10\%$.  

At each time-step, the temperature for the top-most zone is set with respect to that of the zone directly below it according to a power-law solution obtained by assuming a radiative-zero atmosphere above our grid.\footnote{Changing the upper boundary condition to a constant temperature has a negligible effect on the long-term evolution of the layer.  This is unsurprising because, for the pressure at the top of the grid, $P_{\rm top}$, the ratio of the thermal time to the accretion time at the base is $P_{\rm top}/P_b$.  For the grid we use and typical ignition pressures, this ratio is $< 0.1$, and so any differences in thermal conditions at the top of the grid are radiated away and not carried deeper into the star.}  The thermal boundary condition for the bottom-most zone is such that the flux there is equal to that of the zone directly above it.  While locally incorrect, this bottom boundary condition has no effect on the thermal properties of the accreted envelope as the thermal time at the bottom of our grid is much longer than the accretion time prior to ignition.  Thus, any inaccuracies in the bottom-most zones will have no effect on the region of interest.

The WD core structure is assumed to be constant during the nova cycle.  This is an excellent approximation, as the WD's central pressure, $10^{23}-10^{27}$ dyne cm$^{-2}$ for $M = 0.6-1.35 \ M_\odot$, is much larger than that of the accreted layer.  The WD core radius is calculated for an isothermal core that is half $^{12}$C and half $^{16}$O by mass with outer boundary condition $P_b=10^{18}$ dyne cm$^{-2}$.  The radius is relatively insensitive to $T_c$ for $T_c<2\E{7} \ {\rm K}$, so all radii are calculated with $T_c=10^7$ K.  The resulting radii and gravitational accelerations for our 4 models are shown in Table \ref{tab:WDrad}.  Calculations for O/Ne WDs give the same radii to within $\simeq 1\%$, because $^{12}$C, $^{16}$O, and $^{20}$Ne have the same charge-to-mass ratio.

\begin{table}
	\begin{center}
	\caption{WD Core Properties for $T_c=10^7$ K and Outer Boundary Condition $P_b=10^{18}$ dyne cm$^{-2}$}
	\label{tab:WDrad}
	\begin{tabular}{|c|c|c|}
	\hline
	Mass $\left[ M_\odot \right]$ & Radius $\left[ \rm{cm} \right]$ & $g \left[ \rm{cm} \ \rm{s}^{-2} \right]$\\
	\hline
	\hline
	0.6 & $8.39\E{8}$ & $1.13\E{8}$\\
	\hline
	1.0 & $5.40\E{8}$ & $4.55\E{8}$\\
	\hline
	1.2 & $3.83\E{8}$ & $1.09\E{9}$\\
	\hline
	1.35 & $2.26\E{8}$ & $3.52\E{9}$\\
	\hline
	\end{tabular}
	\end{center}
\end{table}

The equation of state \citep{scvh95,ts00,rn02}, opacity \citep{ir93,ir96}, nuclear burning network \citep{tim99}, neutrino cooling \citep{itoh96}, and electron screening \citep{grab73,aj78,itoh79} are calculated using the MESA code package.\footnote{http://mesa.sourceforge.net/}  The MESA basic nuclear network, which tracks the abundances of $^1$H, $^4$He, $^{12}$C, $^{14}$N, $^{16}$O, $^{20}$Ne, and $^{24}$Mg, has been modified to explicitly follow the $^3$He reactions $^3$He$+^3$He and $^3$He$+^4$He.  The nuclear reactions consist of the $p$-$p$ chains (including the $pep$ reaction, although for our high $\dot{M}$'s $\geq 10^{-9} \ \smpy$, it typically has a negligible contribution; \citealt{starr07}), the CNO cycles, $\alpha$-burning up to $^{24}$Mg, and C/O burning.  Isotopes that are not explicitly tracked are assumed to have equilibrium abundances.

\begin{table}
	\begin{center}
	\caption{Solar Photosphere Mass Fractions from \cite{lod03}}
	\label{tab:massfrac}
	\begin{tabular}{|c|c|}
	\hline
	Element & Mass fraction\\
	\hline
	\hline
	$^1$H & 0.749\\
	\hline
	$^4$He & 0.237\\
	\hline
	$^{12}$C & $2.21\E{-3}$\\
	\hline
	$^{14}$N & $0.71\E{-3}$\\
	\hline
	$^{16}$O & $5.87\E{-3}$\\
	\hline
	$^{20}$Ne & $1.12\E{-3}$\\
	\hline
	$^{24}$Mg & $0.64\E{-3}$\\
	\hline
	\end{tabular}
	\end{center}
\end{table}

Solar composition of the accreted material is defined by the recommended solar photosphere abundances in \cite{lod03}.  The relevant elemental abundances by mass fraction are shown in Table \ref{tab:massfrac}; note the reduction in solar metallicity as compared to previous studies, such as \cite{gs98}.  For models with non-solar metallicities, the mass fractions of $^{12}$C, $^{14}$N, $^{16}$O, $^{20}$Ne, and $^{24}$Mg are adjusted by the same multiplicative factor\footnote{As noted in \S\ref{sec:analytics}, systems do exist where the mass fractions of accreted $^{12}$C and $^{14}$N are non-solar while the other metals still have their solar values.  However, for the sake of consistency and convenience, we scale all the metals by the same value in each model.} and the difference made up in $X_{\rm H}$.  Likewise, any non-zero $^3$He mass fraction is subtracted from $X_{\rm H}$.

A monotonic transport first-order advection scheme \citep{vl74,haw84}, modified for logarithmic coordinates, is utilized to simulate accretion.  While certainly more accurate than zeroth-order donor cell advection, it is still subject to numerical advection that smoothes out what should be a step-function accretion front in the absence of diffusion, which we have neglected for the reasons given in \S\ref{sec:motiv}.  If unaccounted for, this non-physical excess advection into C-rich material leads to premature burning of the small amount of hydrogen that precedes the accretion front.  Thus, when calculating the burning rate and compositional changes, the core C/O is treated as $^{24}$Mg, which is essentially inert in the nuclear reaction network at these temperatures.  For all other calculations, the core remains half $^{12}$C and half $^{16}$O by mass.

The code is evolved until any local thermal gradient is steeper than the local adiabatic gradient and convection sets in, at which point we consider ignition to have occurred.  We define the ignition mass to be the total accreted mass at the onset of convection, $M_{\rm ign} = \dot{M} t_{\rm ign}$, taking into account the time required for the accretion front to reach the top of our grid prior to the beginning of the code run.


\subsection{Representative results}

\begin{figure}
	\epsscale{1.0}
	\plotone{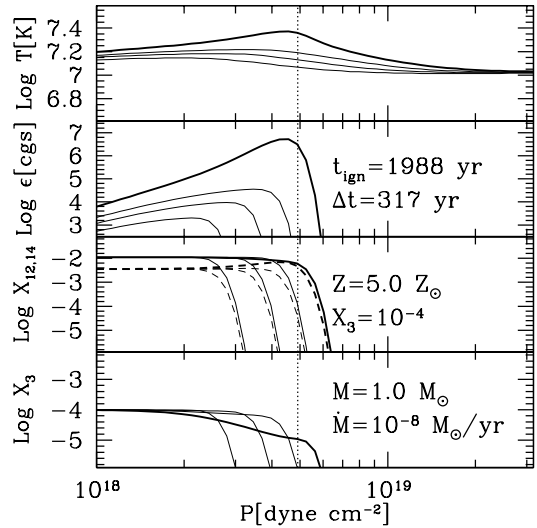}
	\epsscale{1.0}
	\caption{Time series of envelope profiles for a $1.0 \ M_\odot$ WD accreting at $\dot{M}=10^{-8} \ \smpy$ with $Z=5.0 \ Z_\odot$ and $X_3=10^{-4}$. Each profile in each panel is separated by 317 yr.  The thick line is the final profile just prior to convection, 1988 yr after the onset of accretion.  Thus, the first profiles shown represent the layer after 1037 yr of accretion.  From top to bottom, the panels show the temperature, energy generation rate (in cgs units of erg g$^{-1}$ s$^{-1}$), accreted $^{12}$C and $^{14}$N mass fractions, and $^3$He mass fraction.  The solid lines in the third panel show $X_{12}$, and the dashed lines show $X_{14}$.  The vertical dotted line shows the pressure of the envelope base just prior to convection.}
	\label{fig:z5_0}
\end{figure}

\begin{figure}
	\epsscale{1.0}
	\plotone{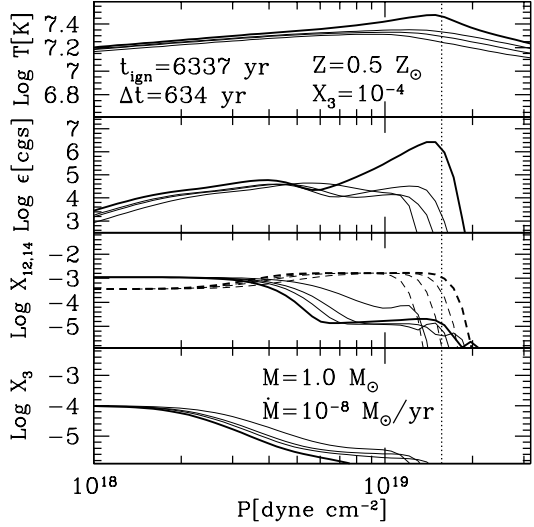}
	\epsscale{1.0}
	\caption{Same as Fig. \ref{fig:z5_0}, except with $Z=0.5 \ Z_\odot$, time between profiles of 634 yr, and $t_{\rm ign}=6337$ yr, so the earliest profiles show the layer after 4434 yr of accretion.}
	\label{fig:z0_5}
\end{figure}

Figures \ref{fig:z5_0} and \ref{fig:z0_5} show envelope profiles for a $1.0 \ M_\odot$ WD accreting at $\dot{M}=10^{-8} \ \smpy$ with $X_3=10^{-4}$.  Figure \ref{fig:z5_0} has $Z=5.0 \ Z_\odot$, and Figure \ref{fig:z0_5} has $Z=0.5 \ Z_\odot$.  Each panel shows profiles at 4 different times, each separated by 317 yr in Figure \ref{fig:z5_0} and by 634 yr in Figure \ref{fig:z0_5}.  The thick solid lines in both figures show the envelope profile just prior to convection.

Figure \ref{fig:z5_0} shows a typical $p+^{12}$C ignition.  The cause of the ignition is the onset of $p+^{12}$C burning, which releases more heat than can be radiatively transported away.  At ignition, the accreted carbon mass fraction (\emph{solid line in the third panel}) at the envelope base (\emph{vertical dotted line}) is essentially the same as in the rest of the accreted layer because it has not had a chance to deplete before ignition conditions are met.  Thus, the $^{14}$N mass fraction (\emph{dashed line in the third panel}) in the layer is also unchanged, except near the base, where the small amount of C-burning has slightly raised the $^{14}$N mass fraction.

Contrast this sequence of events with the $0.5 \ Z_\odot$ case shown in Figure \ref{fig:z0_5}.  Here, carbon has already been depleted prior to ignition.  At the time of ignition, the carbon mass fraction at the base is several orders of magnitude lower than in the rest of the layer (\emph{solid line in the third panel}), and the difference has been added to the $^{14}$N mass fraction (\emph{dashed line in the third panel}).  The energy generation profile (\emph{second panel}) clearly shows two peaks in the envelope: the shallower occurs where $^{12}$C is burned and depleted, and the deeper peak is due to full CNO cycle burning.  It is at the deeper location that $\epsilon_{\rm nuc}$ begins to spike and convection occurs.  Since the layer must become hotter and denser to burn $^{14}$N, more mass accumulates prior to ignition.  The ignition mass in this case is $6.3 \E{-5} \ M_\odot$, \emph{3 times higher than the C-triggered case}, which has $M_{\rm ign}=2.0\E{-5} \ M_\odot$.

\begin{figure}
	\epsscale{1.0}
	\plotone{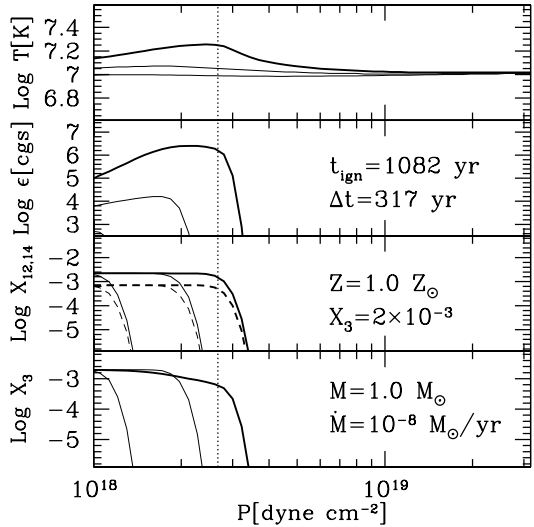}
	\epsscale{1.0}
	\caption{Same as Fig. \ref{fig:z5_0}, except with $Z=Z_\odot$, $X_3=2\E{-3}$, and $t_{\rm ign}=1082$ yr, and thus the earliest time shown is 448 yr after accretion begins.}
	\label{fig:x32_0}
\end{figure}

The disparity in $M_{\rm ign}$ is even greater when compared to a case with a non-negligible amount of $^3$He, because as shown in \S\ref{sec:3heeffect}, $^3$He always begins burning prior to proton captures onto $^{12}$C.  Figure \ref{fig:x32_0} shows the outcome of a model with $X_3=2\E{-3}$.  This ignition is triggered by $^3$He-burning after only $1.1\E{-5} \ M_\odot$ has been accreted.


\subsection{Ignition masses}
\label{sec:mign}

\begin{figure}
	\epsscale{1.0}
	\plotone{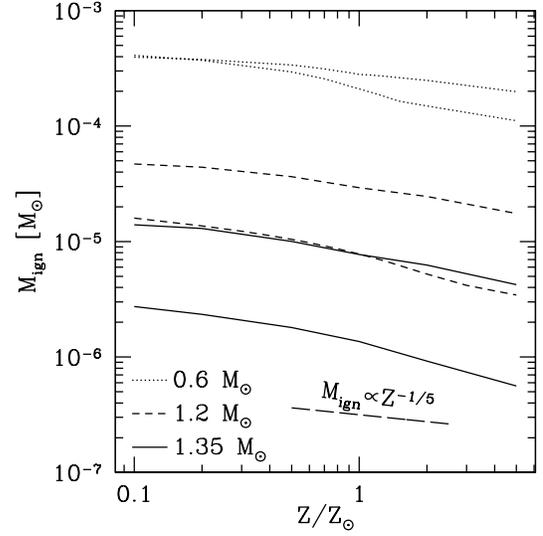}
	\epsscale{1.0}
	\caption{Ignition mass vs. metallicity for $X_3=10^{-4}$ and different WD masses and $\dot{M}$'s.  Dotted, short-dashed, and solid lines are for $M=0.6, \ 1.2, \ {\rm and} \ 1.35 \ M_\odot$ WDs, respectively.  The top (higher $M_{\rm ign}$) curve for each WD mass has $\dot{M}=10^{-9} \ \smpy$, and the bottom (lower $M_{\rm ign}$) curve has $\dot{M}=10^{-8} \ \smpy$ for $M=0.6 \ M_\odot$ and $\dot{M}=10^{-7} \ \smpy$ for $M=1.2$ and $1.35 \ M_\odot$.  The long-dashed line shows the analytic scaling of $M_{\rm ign} \propto Z^{-1/5}$, as discussed in \S\ref{sec:mign}.}
	\label{fig:mignvsz}
\end{figure}

\begin{figure}
	\epsscale{1.0}
	\plotone{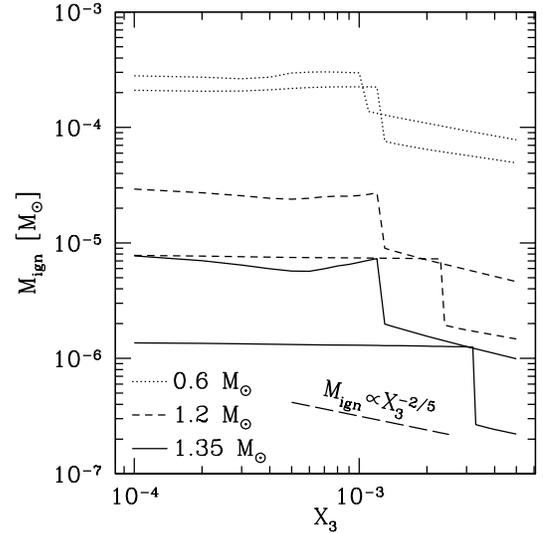}
	\epsscale{1.0}
	\caption{Same as Fig. \ref{fig:mignvsz} except vs. $X_3$.  The metallicity is the solar value.  The analytic scaling from \S\ref{sec:mign} of $M_{\rm ign} \propto X_3^{-2/5}$ is displayed.}
	\label{fig:mignvsx3}
\end{figure}

In this section, we show the resulting $M_{\rm ign}$ for a range of $M$, $\dot{M}$, and composition.  Figure \ref{fig:mignvsz} shows $M_{\rm ign}$ vs. accreted metallicity with $X_3=10^{-4}$, and Figure \ref{fig:mignvsx3} shows $M_{\rm ign}$ vs. $^3$He mass fraction with solar metallicity.  The dotted, dashed, and solid lines represent $M=0.6, \ 1.2, \ {\rm and} \ 1.35 \ M_\odot$ WDs, respectively.  The top (higher $M_{\rm ign}$) curve for each WD mass has $\dot{M}=10^{-9} \ \smpy$.  The bottom (lower $M_{\rm ign}$) curve has $\dot{M}=10^{-8} \ \smpy$ for $M=0.6 \ M_\odot$ and $\dot{M}=10^{-7} \ \smpy$ for $M=1.2$ and $1.35 \ M_\odot$.  The highest $\dot{M}$ shown for $M=0.6 \ M_\odot$ differs from that of the other masses because a $0.6 \ M_\odot$ WD accreting at $10^{-7} \ \smpy$ burns hydrogen stably without any novae \citep{sien75,sien80,pac83,fuji82b,sb07,nom07}.

Some general trends are clear.  Increasing the WD mass and accretion rate lower the ignition mass.  A larger mass is equivalent to a smaller radius, both of which contribute to a higher value of $g$, resulting in a higher temperature and density for a given envelope mass, and thus ignition conditions are reached at lower envelope masses.  Higher accretion rates translate into higher compressional luminosities and higher temperatures, also resulting in ignition for smaller envelope masses.  The basic trend of lower ignition mass with higher values of $^{12}$C and $^3$He is also sensible: more fuel means more burning and quicker buildup to ignition.

However, as we have discussed previously, changing the composition does not just lead to a change in the energy generation rate.  If the $^3$He or $^{12}$C is depleted, conditions for burning the next fuel will have to be reached for ignition.  This is the cause of the inflections that are present in most of the $M_{\rm ign}$-curves in Figure \ref{fig:mignvsz}, and for the abrupt changes in $M_{\rm ign}$ that are seen in Figure \ref{fig:mignvsx3}.  Once a critical $^{12}$C- or $^3$He-mass fraction is reached, the ignition becomes qualitatively different.  Note that carbon depletion does not have as drastic an effect as $^3$He depletion because depleted carbon is converted to another burning catalyst, $^{14}$N, whereas $^3$He is depleted to hydrogen and $^4$He, negligibly increasing the amount of hydrogen fuel.

The long-dashed lines in Figures \ref{fig:mignvsz} and \ref{fig:mignvsx3} show the analytic power-law scaling of $M_{\rm ign}$ with $Z$ and $X_3$ as derived from equation (\ref{eq:unstab}).  Assuming that the constant pressure logarithmic temperature dependences of the $^{12}$C- and $^3$He-burning rates are $\chi_{12} \simeq \chi_3 \simeq 15$ yields $M_{\rm ign} \propto Z^{-1/5}$ for $^{12}$C-triggered novae and $M_{\rm ign} \propto X_3^{-2/5}$ for $^3$He-triggered novae.  The difference in the exponent is due to $^3$He burning via self-reactions, as opposed to $^{12}$C burning via proton captures.  The numerical results of Figure \ref{fig:mignvsz} match the analytic scaling fairly well.  The high $X_3$ models of Figure \ref{fig:mignvsx3} also match the analytic expectation well.  However, for lower values of $X_3$, $^3$He is depleted and does not release enough heat to trigger the nova.  These novae are instead triggered by $^{12}$C or $^{14}$N, and since the metallicity is constant along each curve, $M_{\rm ign}$ is also roughly constant below a critical value of $X_3$.

\begin{figure}
	\epsscale{1.0}
	\plotone{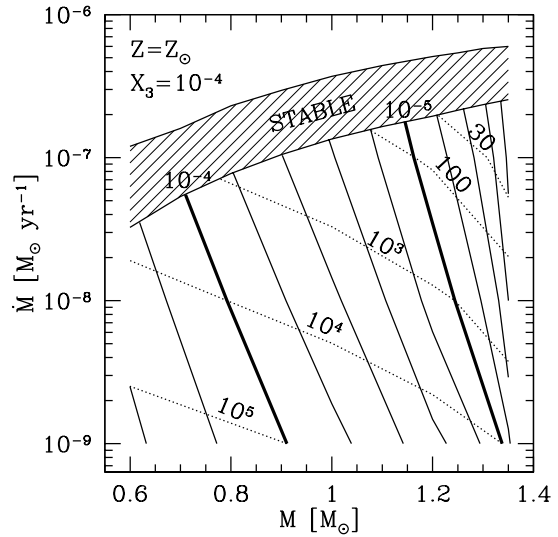}
	\epsscale{1.0}
	\caption{Lines of constant $M_{\rm ign}$ in $M-\dot{M}$ space for $Z=Z_\odot$ and $X_3=10^{-4}$ (\emph{solid lines}).  Contours are evenly spaced with 0.2 decades of $M_{\rm ign}$ between each line.  Thick solid lines have $M_{\rm ign}=10^{-5}$ and $10^{-4} \ M_\odot$ as labeled.  Also shown are lines of constant recurrence time (\emph{dotted lines}), labeled in years.  The range of stable H-burning (\emph{hatched region}) is obtained from \cite{nom07}.}
	\label{fig:migncont}
\end{figure}

Figure \ref{fig:migncont} shows lines of constant $M_{\rm ign}$ as a function of $M$ and $\dot{M}$ (\emph{solid lines}) for $Z=Z_\odot$ and $X_3=10^{-4}$, obtained by quadratically fitting the numerical results detailed above.  The contours are evenly spaced with 0.2 dex of $M_{\rm ign}$ between each line.  The thick solid lines have $M_{\rm ign}=10^{-5}$ and $10^{-4} \ M_\odot$ as labeled.  Also shown are contours of constant recurrence time (\emph{dotted lines}), labeled in years.  The hatched region where hydrogen burns stably in a steady state is taken from \cite{nom07}.  It is clear that a recurrent nova system accreting solar metallicity material with low $^3$He mass fraction must have a WD $> 1.2 \ M_\odot$ to achieve a recurrence time $<30$ yr.  

Our ignition masses are a factor of $1.5-3$ times higher than the analytic estimates of \cite{fuji82b} and $1.1-2$ times lower than the work of \cite{mac83}.  Given the differences in method and updates in equations of state and opacity, these discrepancies are not significant.  The work of \cite{yaron05} assumes diffusive accretion onto initially bare C/O cores, which invalidates our assumption of accretion onto C-poor material, so the fact that their $M_{\rm ign}$'s are $2-4$ times lower than ours is not surprising.  A comparison of our $X_3=10^{-3}$ results with those of \cite{tb04} (and Fig. 1 of \citealt{tb05}) reveals a large discrepancy, with our $M_{\rm ign}$'s higher by a factor of $3-10$.  The difference is predominantly due to their lack of abundance evolution, which eliminates the possibility of $^3$He depletion.  Theirs is a valid assumption for  higher mass fractions $X_3 \gtrsim 2\E{-3}$, but is not correct for $X_3=10^{-3}$, and leads to premature nova ignitions.

Throughout this study, we have assumed that $M_{\rm ign}$ and $t_{\rm ign}$ are the accreted mass and accretion time prior to the onset of convection.  However, the convective zone takes some time to grow and reach the surface, at which point the nova outburst becomes observable.  It is conceivable that a significant amount of mass is accreted during the convective phase, which would cause the $t_{\rm ign}$ we have calculated to be lower than the actual time between novae.  To quantify this error, we consider the growth time of the convective zone, which is $t_{\rm growth} \sim c_P T_b / \epsilon$, where the temperature and burning rate are evaluated at the onset of convection.  For each model, we find that $t_{\rm growth}$ is always $20-100$ times shorter than the time to the onset of convection, so that the ignition times and masses we have calculated are only smaller than the actual values by a few percent, and our assumption is justified.


\subsection{Ignition pressures and the misuse of ``$P_{\rm crit}$''}

\begin{figure}
	\epsscale{1.0}
	\plotone{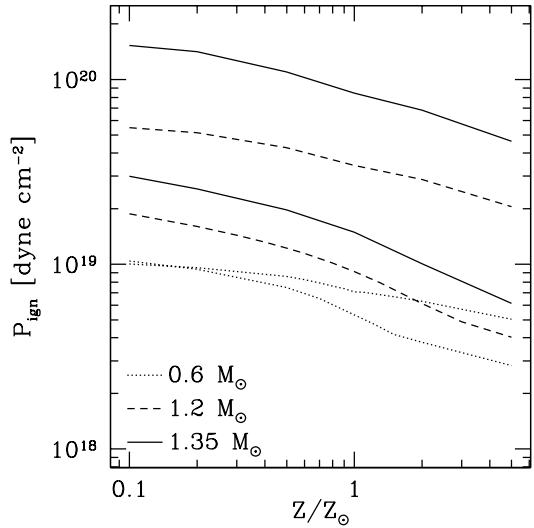}
	\epsscale{1.0}
	\caption{Ignition pressure vs. metallicity for $X_3=10^{-4}$ and a range of $M$ and $\dot{M}$.  Dotted, dashed, and solid lines are for $M=0.6$, $1.2$, and $1.35 \ M_\odot$, respectively.  Lower $P_{\rm ign}$ curves for each WD mass have $\dot{M}=10^{-9} \ \smpy$.  The higher $P_{\rm ign}$ curve for $M=0.6 \ M_\odot$ has $\dot{M} = 10^{-8} \ \smpy$, and the higher $P_{\rm ign}$ curves for $M=1.2$ and $1.35 \ M_\odot$ have $10^{-7} \ \smpy$.}
	\label{fig:pignvsz}
\end{figure}

\begin{figure}
	\epsscale{1.0}
	\plotone{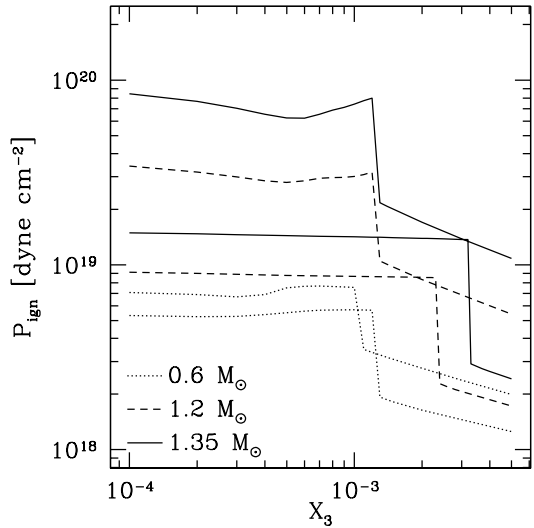}
	\epsscale{1.0}
	\caption{Same as Fig. \ref{fig:pignvsz}, except vs. $X_3$ with solar metallicity.}
	\label{fig:pignvsx3}
\end{figure}

\cite{fuji82a} is often cited to support the claim of a critical ignition pressure for nova ignition, $P_{\rm crit}$, which, depending on the study making the claim, has a value between $10^{19}-10^{20}$ dyne cm$^{-2}$.  However, the ignition pressure is clearly not constant, as we show in Figures \ref{fig:pignvsz} and \ref{fig:pignvsx3}.  These figures are identical to Figures \ref{fig:mignvsz} and \ref{fig:mignvsx3} but with $P_{\rm ign}$ along the ordinate axis instead of $M_{\rm ign}$; the ignition pressure is related to $M_{\rm ign}$ through equation (\ref{eq:thinshell}).  The ignition pressure varies by a factor of 100 from one extreme of high accretion rate, mass, and $X_3$ to the other extreme and is not constant.  Moreover, the original paper actually makes no such assertion.  Instead, \cite{fuji82a} states that there is a minimum ignition pressure necessary to produce a strong nova-like outburst powered by hydrostatic shell expansion.  The ignition pressure is left as a free parameter and is calculated in \cite{fuji82b} as a function of WD masses and accretion rates.


\subsection{Pre-ignition luminosities}
\label{sec:obs}

The depletion of fuel will increase the WD surface luminosity above the compressional luminosity.  Such an increase in the pre-nova light curve would be a sign that the carbon abundance is not enhanced, or that the $^3$He abundance is non-negligible.  There are several energy scales to keep in mind: the accretion, thermal, $^{12}$C-burning, and $^3$He-burning energies per mass are given, respectively, as
\be
	E_{\rm acc} = \frac{GM}{R} &=& 1.3\E{17} \ \frac{M_1}{R_9} \frac{\rm erg}{\rm g} \nonumber \\
	E_{\rm therm} = \frac{kT}{\mu m_p} &=& 1.4\E{15} \ T_7 \frac{\rm erg}{\rm g} \nonumber \\
	X_{12} E_{12} &=& 8.8\E{14} \ \frac{X_{12}}{10^{-3}} \frac{\rm erg}{\rm g} \nonumber \\
	X_3 E_3 &=& 2.1\E{15} \ \frac{X_3}{10^{-3}} \frac{\rm erg}{\rm g} .
\ee
The luminosities associated with these energies can be obtained by multiplying them by the accretion rate, with an additional factor $f=1.75$ for the thermal/compressional luminosity.  The radius is typically a few$\E{8}$ cm, so the accretion energy is roughly two orders of magnitude larger than the thermal or nuclear energy available during the accretion phase.  However, the accretion luminosity from a disk is variable and greatly reduced in disk quiescence.  Thus, it is still possible to observe the luminosity produced from the interior of the accreted envelope, as demonstrated by \cite{tb03} in their work on relating effective temperature, $T_{\rm eff}$, to $\dot{M}$.

\begin{figure}
	\epsscale{1.0}
	\plotone{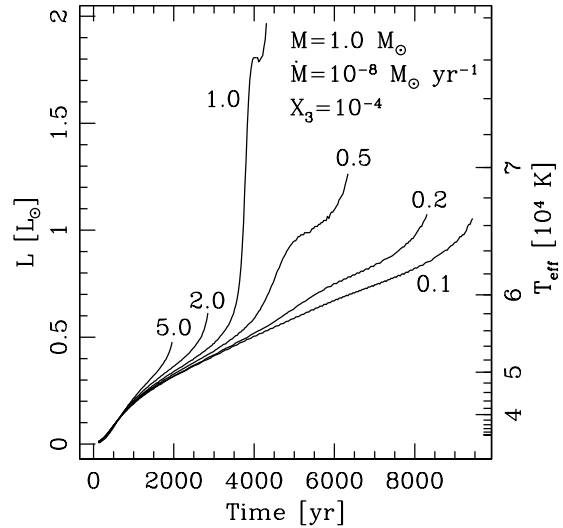}
	\epsscale{1.0}
	\caption{Surface luminosity, excluding the accretion luminosity, and effective temperature as a function of time for a $1.0 \ M_\odot$ WD with $\dot{M}=10^{-8} \ \smpy$, $X_3=10^{-4}$, and various metallicities.  The metallicities shown are 0.1, 0.2, 0.5, 1.0, 2.0, \& 5.0 $Z_\odot$, increasing from right to left.}
	\label{fig:lumz}
\end{figure}

Figure \ref{fig:lumz} shows the surface luminosity, ignoring the accretion luminosity, and effective temperature as a function of time for a $1.0 \ M_\odot $ WD with $ \dot{M}=10^{-8} \ \smpy$, $X_3=10^{-4}$, and metallicities of 0.1, 0.2, 0.5, 1.0, 2.0, \& 5.0 $Z_\odot$, increasing from right to left.  The 0.5 $Z_\odot$ model is interesting in that there is enough carbon to power significant nuclear luminosity, but not enough to trigger a nova.  For this case, the compressional luminosity is $<0.5 \ L_\odot$ and $T_{\rm eff}<5.5\E{4}$ K for $\sim50\%$ of the nova cycle.  Four thousand years after the onset of accretion, carbon is burned and depleted, causing the luminosity from inside the envelope to double within a span of only 1000 yr.  The burning of $^{14}$N begins $\sim 1000$ yr after C-depletion, and the luminosity and effective temperature rise to $1.3 \ L_\odot$ and $8\E{4}$ K just prior to the CNO cycle-triggered ignition.

\begin{figure}
	\epsscale{1.0}
	\plotone{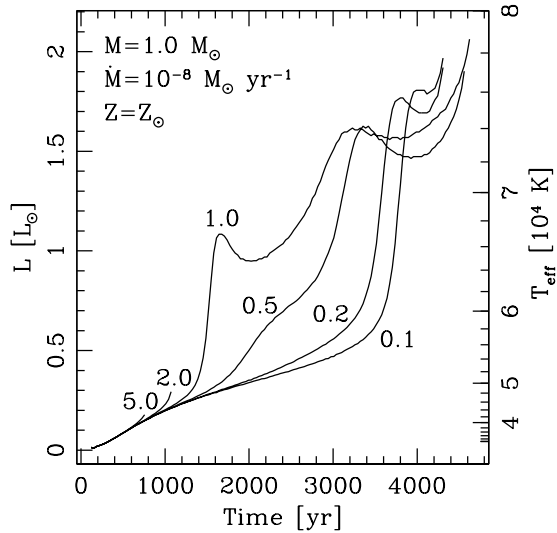}
	\epsscale{1.0}
	\caption{Same as Fig. \ref{fig:lumz}, but with $Z=Z_\odot$ and varying $X_3$.  The $^3$He mass fractions shown are 0.1, 0.2, 0.5, 1.0, 2.0, \& $5.0\E{-3}$, increasing from right to left.  }
	\label{fig:lumx3}
\end{figure}

Changing $X_3$ also has an effect on the surface luminosity and $T_{\rm eff}$.  Figure \ref{fig:lumx3} shows the surface luminosity and effective temperature as in Figure \ref{fig:lumz}, but with fixed metallicity $Z=Z_\odot$ and varying $^3$He mass fractions of 0.1, 0.2, 0.5, 1.0, 2.0, \& $5.0\E{-3}$, increasing from right to left.  Here, the interesting case is $X_3=10^{-3}$, which has enough $^3$He to produce significant energy, but not enough to trigger the nova.  The surface luminosity reaches $L_\odot$ and $T_{\rm eff}$ rises to $6.5\E{4}$ K after only 1500 yr of accretion, but the surface then cools as the envelope succeeds in depleting the accumulated $^3$He.  The surface brightens when C-burning commences, and again dims when carbon is depleted, finally rising to $2 \ L_\odot$ and $8\E{4}$ K when $^{14}$N begins burning.


\section{Conclusions}
\label{sec:conc}

Motivated by uncertainties in classical nova core-mixing mechanisms and the lack of metal enhancements in some nova ejecta, we have quantified the effects of composition on nova ignitions (see Fig. \ref{fig:mignvsz}) under the assumption that the underlying material is C-poor and diffusion thus unimportant, as appropriate for accretors with large helium buffers or O/Ne cores.  We have found that for carbon mass fractions $\lesssim 2\E{-3}$, $^{12}$C is depleted and converted to $^{14}$N without releasing enough heat to trigger a nuclear instability.  The layer continues to accrete until $^{14}$N can capture protons, leading to a nova triggered by the full CNO cycle and an ignition mass larger than the carbon-ignited case.  Thus, the ignition mass increases by a factor of $\sim 3$ as the metallicity is decreased from $5.0 \ Z_\odot$ to $0.1 \ Z_\odot$.  The critical carbon mass fraction is near-solar and is thus relevant to sub-solar metallicity systems as well as systems with evolved secondaries that have undergone CNO processing of $^{12}$C to $^{14}$N.  We have also examined the effect of accreted $^3$He (see Fig. \ref{fig:mignvsx3}), which can reach mass fractions of $4\E{-3}$ as an evolved donor's interior is uncovered by mass transfer.  For $X_3 \gtrsim 2\E{-3}$, $^3$He+$^3$He reactions trigger novae with $M_{\rm ign} $ a factor of $ \sim 3 $ times smaller than the C-triggered case.

The dependence of $M_{\rm ign}$ on accreted composition will affect population-averaged nova rates: naively, high-metallicity environments would have nova rates higher by a factor of $\sim 3$ than systems with sub-solar metallicities (such as novae in globular clusters; \citealt{sq07}) or evolved donors that have undergone CNO processing.  However, the existence of $^3$He would have to be taken into account, because old systems with donors that have undergone significant mass loss could have $X_3\gtrsim2\E{-3}$ and would thus have high nova rates, regardless of the accreted metallicity.  A proper prediction of the effect of donor composition requires a population-synthesis calculation that includes further complications such as binary and donor evolution.  We leave this exercise for future work.

To date, most observations only report galactically-averaged nova rates, although some M31 studies \citep{ciar87,cap89,darn06} have found that M31's bulge produces more novae per stellar luminosity than its disk by a factor of $\sim 5$.  On the other hand, galactically-averaged nova rates do not see any morphology dependence, finding instead that the luminosity-specific nova rate (LSNR) is roughly constant across all galaxy types at a value of $2 \pm 1$ yr$^{-1}$ $(10^{10} \ L_{\odot, \, K})^{-1}$ \citep{ws04}, where $L_{\odot, \, K}$ is the \emph{K}-band solar luminosity.  The LMC and SMC (and possibly Virgo dwarf elliptical galaxies; \citealt{ns05}), which have LSNRs higher by a factor of 3, are exceptions.  These measurements have large error bars due to small number statistics and issues of completion caused by both extinction and infrequent observations.  A better measurement of nova rates will come with new deep optical surveys with high cadences such as Pan-STARRS-1, Pan-STARRS-4, and the Large Synoptic Survey Telescope, which will see thousands of novae every year.  These will reduce the nova rate error bars and also possibly allow us to measure rates in different populations within other galaxies besides M31.

In addition to this population-averaged observable, the composition could also have a detectable effect on individual systems.  In particular, the depletion of fuel can significantly increase the surface luminosity above the baseline set by the entropy released during compression of the accreted layer (see Figs. \ref{fig:lumz} and \ref{fig:lumx3}).  While this increase is still well below the accretion luminosity associated with gravitational energy release, it would be visible while the system was in disk quiescence.  Recurrent novae, in particular, would be ideal systems in which to observe this brightening in quiescent luminosity due to their short recurrence times of $\sim 30 $ yr.

These observables are dependent on the assumption that convection is not initiated in C-enhanced material.  If instead H-rich envelope material penetrates into C-rich material and triggers convection there, the accreted composition will have little effect.  Thus, if these composition-dependent effects are observed, they will provide evidence that convection for many novae is initiated in C-poor material, and that CNO enrichment for these novae is due to convective shear mixing or overshoot.


\acknowledgments
 
We thank  D. Townsley for discussions, B. Paxton and F. Timmes for valuable assistance with MESA, and J. Steinfadt for nova rate calculations.  This work was supported by the National Science Foundation under grants PHY 05-51164 and AST 07-07633.



\end{document}